\documentstyle[12pt]{article}
\def\e{{\rm e}}
\newcommand{\be}{\begin{equation}}
\newcommand{\ee}{\end{equation}}
\newcommand{\bea}{\begin{eqnarray}}
\newcommand{\eea}{\end{eqnarray}}
\newcommand{\nn}{\nonumber}
\renewcommand{\thefootnote}{\fnsymbol{footnote}}
\begin{document}
\renewenvironment{thebibliography}[1]
  { \begin{list}{\arabic{enumi}.}
    {\usecounter{enumi} \setlength{\parsep}{0pt}
     \setlength{\itemsep}{3pt} \settowidth{\labelwidth}{#1.}
     \sloppy
    }}{\end{list}}
\parindent=1.5pc
\begin{titlepage}
\rightline{hep-th/9410178}
\rightline{MPI-PhT/94-65}
\rightline{ZU-TH 31/94}
\rightline{October 1994}

\vglue 1.0cm
\begin{center}{{\bf FERMIONS IN THE BACKGROUND OF \\
 \vglue 10pt
                         DILATONIC SPHALERONS
               }\\
\vglue 5pt
\vglue 1.0cm
{GEORGE LAVRELASHVILI
\footnote{Electronic address: lavrela@physik.unizh.ch}
 \footnote{On leave of absence from Tbilisi
Mathematical Institute, 380093 Tbilisi, Georgia}
 }\\
\baselineskip=14pt
\vglue 0.3cm
{\it Institut f\"ur Theoretische Physik, Universit\"at Z\"urich}\\
{\it Winterthurerstrasse 190, CH-8057 Z\"urich, Switzerland
\footnote{Address after October 1, 1994.}
}\\
 \vglue 0.3cm {and}\\
\vglue 0.3cm
{\it Max-Planck-Institut f\"ur Physik, Werner-Heisenberg-Institut}\\
{\it F\"ohringer Ring 6, D-80805 Munich, Germany}
\vglue 0.8cm
{ABSTRACT}}
\end{center}
\vglue 0.3cm
{\rightskip=3pc
 \leftskip=3pc
 \tenrm\baselineskip=12pt
\noindent
We discuss the properties and interpretation of a discrete sequence
of a static spherically symmetric solutions of the Yang-Mills-dilaton
theory. This sequence is parametrized by the number $n$ of zeros of a
component of the gauge field potential. It is demonstrated that
solutions with odd $n$ posses all the properties of the sphaleron.
It is shown that there are normalizable fermion zero modes in the
background of these solutions. The question of instability is
critically analysed.
\vglue 0.8cm}
\end{titlepage}
\setcounter{footnote}{0}
\renewcommand{\thefootnote}{\arabic{footnote}}
{\bf\noindent 1. Introduction}
\vglue 0.4cm
\baselineskip=14pt
\noindent
A discrete sequence of static finite energy solutions of a
Yang-Mills-dilaton (YMD) theory was found recently
\cite{LM1} (see also \cite{BIZ1}).
These solutions can be labeled by the number of zeros of the
gauge function and have the same nature as the Bartnik-McKinnon (BMK)
solutions \cite{BMK} of the  coupled Einstein-Yang-Mills (EYM)
equations. On the other hand both the BMK solutions and the solutions
of the YMD theory are related to usual electroweak
sphaleron \cite{DHN}, \cite{MAN}, \cite{KMAN}, \cite{VG}.

The sphaleron solution was found in ref. \cite{DHN} and
later was interpreted \cite{MAN}, \cite{KMAN} as
a solution, which lies on the top of the energy barrier
separating topologically different vacua \cite{JRCDG}.
Creation and subsequent decay of the sphaleron leads to the
fermion number non-conservation \cite{THOOFT} related to the existence
of the Adler-Bell-Jackiw (ABJ) anomaly \cite{ABJ}.
This fermion number non-conservation has natural interpretation
in terms of the phenomenon of fermion level crossing
(see e.g., \cite{MRTS} and ref. therein)
in the background of a topologically nontrivial gauge field.
The observation is that a fermion zero mode occurs,
when the gauge field passes the sphaleron configuration.

The solutions of YMD theory
with an odd number of zeros of the gauge function carry all
the main properties of the sphaleron, namely:\\
(i) they have finite energy\\
(ii) they are saddle points of the action functional\\
(iii) they have fractional topological charge \\
(iv) there are fermion zero modes in background of these
solutions.\\
Therefore it is natural to interpret them as a dilatonic sphalerons.

In the present paper we will critically review the stability analysis
and demonstrate  properties (iii) and (iv) of the dilatonic sphalerons.

The rest of the paper is organized as follows.
In the next section we will review  dilatonic sphaleron solutions.
In section 3 we will show that similarly as for the electroweak
sphaleron one can assign a (half-integer) topological charge
to the  dilatonic sphalerons.
In section 4 we will demonstrate existence of the fermion zero
modes in the background of the dilatonic sphalerons.
In section 5 we will present the stability analysis.
Section 6 contains concluding remarks.

\vglue 0.6cm
{\bf\noindent 2. Dilatonic Sphalerons \hfil}
\vglue 0.2cm
We are interested in YMD theory defined by the action
\be
S=
\frac{1}{4\pi}\int d^4x\Bigl(
\frac{1}{2}(\partial_{\mu}\varphi)(\partial^{\mu}\varphi)-
{\e^{2\kappa\varphi}\over 4 g^2 }F^a_{\mu\nu} F^{a~\mu\nu}
 \Bigr) \label{act4}
\ee
where $F^a_{\mu\nu}$ is a $SU(2)$ gauge field strength
$F^a_{\mu\nu}=\partial_{\mu} W^a_{\nu}-
\partial_{\nu} W^a_{\mu}+
\epsilon^{abc} W^b_{\mu}W^c_{\nu}$
and $a=1,2,3$ is $SU(2)$ group index, $\mu,\nu = 0,1,2,3$
are space-time indexes.
\footnote{Our conventions are as follows:
  $A_\mu B^\mu = A_0 B^0 + A_i B^i =
  A_0 B_0 - A_i B_i$ where $i,j=1,2,3$ are space indexes.}
Using scaling properties of the action Eq.(\ref{act4}) one
can remove the dependence on the dilatonic coupling constant
$\kappa$ and the gauge coupling constant $g$ from $S$,
therefore we may put $\kappa=g=1$ in what follows without restriction.

We are interested in spherically symmetric solutions.
For our analysis we will essentially adopt the notations of
refs. \cite{BFM}, \cite{LM3}.
The most general spherically symmetric
ansatz for the {\sl SU(2)\/} Yang-Mills field $W_\mu^a$ can be written
(in the Abelian gauge) as \cite{ANSATZ}
\bea
 W_t^a&=(0,0,A_0)\,,\qquad  W_\theta^a&=(\phi_1,\phi_2,0)\, \nn \\
 W_r^a&=(0,0,A_1)\,, \qquad
 W_\varphi^a&=(-\phi_2 \sin\theta,\phi_1\sin\theta,\cos\theta)\,.
\label{gauge}
\eea
This ansatz Eq.(\ref{gauge}) is form invariant
under gauge transformations around the third isoaxis,
with $A_\alpha$ transforming as a $U(1)$ gauge field
on the reduced space-time $(t, r)$, whereas
$\phi=\phi_1+i\phi_2$ is a scalar of charge one with the
covariant derivative
$D_\alpha\phi=\partial_\alpha\phi-iA_\alpha\phi$.
With respect to this $U(1)$ one may define the `charge conjugation'
$\phi\to\overline\phi$, $A_\alpha\to -A_\alpha$.
The corresponding YMD action is
\be
S=\int drdt\Bigl[
{r^2\over 2}
(\partial_{\alpha}\varphi)(\partial^{\alpha}\varphi)
-\e^{2\varphi}\Bigl(
{r^2\over4}F^{\alpha\beta}F_{\alpha\beta}
-\overline{D^\alpha\phi} D_\alpha\phi+
{1\over2r^2}(|\phi|^2-1)^2
\Bigr)\Bigr]\;,
\label{actYMD}
\ee
where $F_{\alpha\beta}$ denotes the field strength of $A_\alpha$
, ($\alpha, \beta = t, r$).

Choosing the gauge $A_0=0$ the YMD action  Eq.(\ref{actYMD}) can
be written more explicitly
\be
S=\int drdt\Bigl[{r^2\over 2}\dot{\varphi}^2
-{r^2\over 2}\varphi'^2
-\e^{2\varphi}\Bigl(-{r^2\over 2}\dot A_1^2-
|\dot\phi|^2+|\phi'-iA_1\phi|^2+{1\over2r^2}(|\phi|^2-1)^2
\Bigr)\Bigr]\;,
\label{act1}
\ee
where a prime denotes $d\over dr$ and a dot $d\over dt$.

It was found \cite{LM1}, \cite{BIZ1}  that the YMD theory
in the even sector with respect to $U(1)$ charge conjugation,
\be
A_0 =0,\; A_1 =0,\; \phi_1 \equiv W(r),\; \phi_2 =0,
\label{ansatz2}
\ee
 has a discrete sequence of static solutions.
In this sector the ansatz Eq.(\ref{gauge}) is equivalent
to the usual ``monopole"  ansatz
\be
W_0^a=0, \quad
W_i^a=\epsilon_{aij}{n_j\over r}(1-W(r))
\label{ansatz}
\ee
with  $n_j=x_j/r$.

The corresponding reduced action is
\be
S^{\rm red}=-\int dr\Bigl[{r^2\over 2}\varphi'^2+
\e^{2\varphi}\Bigl(W'^2+{(W^2-1)^2\over 2r^2}\Bigr)\Bigr]\,
\ee

The resulting field equations are
\bea
  W''&=&{W(W^2-1)\over r^2}
              -2\varphi' W'\;, \nn \\
 (r^2\varphi')'&=&
 2\e^{2\varphi}\Bigl(W'^2+{(1-W^2)^2\over2r^2}
 \Bigr)\;.\label{eqm}
\eea
These equations are invariant under a shift
$\varphi\to\varphi+\varphi_0$ accompanied by a simultaneous rescaling
$r\to r\e^{\varphi_0}$, hence
globally regular solutions can be normalized to $\varphi(\infty)=0$.

It was found \cite{LM1}, \cite{BIZ1} that
Eqs.(\ref{eqm}) have a discrete sequence of finite energy
solutions $\{W_n, \varphi_n\}$, where $n=1,2,3,...$
labels number of zeros of the gauge function $W_n(r)$.

The mass of this solution varies from $\approx 0.8$ for $n=1$
to $1.0$ for $n\to\infty$ in natural units $1/\kappa g$.

The asymptotic behavior of these solutions for $r\to 0$ is
\bea
W_{n}(r)&=&1-b_{n}r^2+O(r^4)\;, \nn \\
\varphi_{n}(r)&=&\varphi_{n}(0)+
2\e^{2\varphi_{n}(0) }b_{n}^2r^2+O(r^4)\;,\label{ezero}
\eea
and for $r\to \infty$ is
\bea
W_{n}(r)&=&(-1)^{n}(1-{c_{n}\over r}+O({1\over r^2}))\;, \nn \\
\varphi_{n}(r)&=&-{d_{n}\over r}+O({1\over r^4})\;.\label{einfty}
\eea

The parameters $b_n,\varphi_n(0), c_n$ and $d_n$
have to be determined from the numerical calculations.

Note that at infinity $r \to \infty$ these solutions are pure
gauges (compare Eq.(\ref{ansatz}))
\be
\frac{\tau^a}{2}W^a_j(\vec{x})\equiv W_j(\vec{x})=
-\frac{i} {g}(\partial_j U)U^{-1}\;, \label{gauges}
\ee
with
\be
U=\e^{-\frac{i} {2}\pi\vec{\tau}\cdot \vec{n}}
\label{gaugeu}
\ee
for $n$ odd and $U=1$ for even $n$.

\vglue 0.6cm
{\bf\noindent 3. Fractional charge  \hfil}
\vglue 0.4cm
One can assign a fermion (baryon) number to the dilatonic sphaleron
exactly in the same way as for the electroweak sphaleron
\cite{KMAN}. It comes out to be $\frac{1}{2}$.

The argument goes as follows. Because of the ABJ anomaly \cite{ABJ}
the fermion current is not conserved at the quantum level
\be
\partial_{\mu} j^{\mu}_B=\frac{1}{32\pi^2}
F^a_{\mu\nu}\tilde F^{a~\mu\nu}\;, \label{anomaly}
\ee
where $\tilde F^{a}_{\mu\nu}=\frac1 2
\epsilon_{\mu\nu\rho\sigma}F^{a~\rho\sigma}$.
Change in baryon charge $Q_B=\int d^3 x j^0_B$ is related
to the topological properties of
the gauge field configuration.
Integrating Eq.(\ref{anomaly}) and
neglecting a possible (integer) contribution from a baryon
current at infinity one obtains
\be
Q_{B}(t_2)-Q_{B}(t_1)=\frac{g^2}{32\pi^2}\int_{t_1}^{t_2}
 dt\int d^3\! x F^a_{\mu\nu}\tilde F^{a~\mu\nu}
\ee
It is well known that $F\tilde F$ is a total divergence
$F^a_{\mu\nu}\tilde F^{a~\mu\nu}=\partial_\mu K^\mu$
where
\be
K^\mu =\epsilon^{\mu\nu\rho\sigma}
(F^a_{\nu\rho}W^a_{\sigma}-\frac{1}{3}g
\epsilon_{abc}W^{a}_{\nu}W^b_{\rho}W^c_{\sigma})\;.
\ee
Using this fact one can write
\be
Q_{B}(t_2)-Q_{B}(t_1)=N_{CS}(t_2)-N_{CS}(t_1)+
\frac{g^2}{32\pi^2}\int_{t_1}^{t_2}dt
\int_{\it S_\infty}d\vec{s}\cdot \vec{K}
\ee
where ${\it S_\infty}$ is the sphere at spatial infinity and
\be
N_{CS}(t)=
\frac{g^2}{32\pi^2}\int_{x^{0}=t}d^3 x\; K^0
\ee
is the Chern-Simons number of the field at the time $t$.

Now we can consider a time dependent gauge field starting at
$t=t_1$ at the trivial vacuum (with $N_{CS}=0$) and arriving
at $t=t_2$ at the sphaleron configuration. Assuming that
$Q_{B}(t_1)=0$ we obtain
\be
Q_{B}(sphaleron)=N_{CS}(t_2)+
\frac{g^2}{32\pi^2}\int_{t_1}^{t_2}dt
\int_{\it S_\infty}d\vec{s}\cdot \vec{K}\;.\label{qsph}
\ee
In the radial gauge $n_{i}W_{i}=0$ the pure gauge configuration
Eqs.(\ref{gauges}), (\ref{gaugeu}), which the sphaleron
(with odd $n$) approaches for $r \to \infty$
is direction dependent. In this gauge $Q_B$ picks
up its value from the second term in Eq.(\ref{qsph}).
Since $Q_B$ is gauge invariant (up to large gauge transformations)
one can choose a more convenient gauge in which the gauge potential
$W_j$ vanishes faster than
$\frac{1}{r}$ and $Q_B$ has no contribution from
the second term in Eq.(\ref{qsph}). For this purpose one
has to apply a gauge transformation $\Omega$, which is regular
at the origin $r=0$ and removes the direction dependence
of $W_j$ at infinity. A possible choice for $\Omega$ is
\be
\Omega=\e^{\frac{1}{2}i\Theta (r)\vec{\tau}\cdot \vec{n}}
\ee
where $\Theta\in[0,\pi]$. Under this gauge transformation the gauge
configuration Eq.(\ref{ansatz}) goes to
\be
W^{a}_{j}\to\tilde W^{a}_{j}=\frac{Wcos\Theta -1}{gr}
\epsilon_{jak}n_k+\frac{W sin\Theta}{gr}(\delta_{aj}-n_{a}n_{j})
+\frac{1}{g}\frac{d\Theta}{dr}n_{a}n_{j}\;. \label{tr}
\ee
If $\Theta\to\pi$ fast enough for $r\to\infty$ the potential
Eq.(\ref{tr}) vanishes faster then $\frac{1}{r}$ and no contribution to
$Q_B$ from the surface term Eq.(\ref{qsph}) arises. Substitution
of Eq.(\ref{tr}) into Eq.(\ref{qsph}) gives
\be
Q_B=\frac{1}{2\pi}[\Theta (r)-W sin\Theta (r)]_{0}^{\infty}
=\frac{1}{2}.
\ee
Thus the solution with odd $n$ have $Q=\frac{1}{2}$ and the
solutions with even $n$ $Q=0$.
\vglue 0.6cm
{\bf\noindent 4. Fermion  Zero Mode \hfil}
\vglue 0.2cm
In this section we will analyse the existence of fermion
zero modes in the background of the dilatonic sphalerons.
The situation is very similar to the one considered in the
paper by Gibbons and Steif \cite{GS}, who analysed the same
problem for the BMK solutions.

The Dirac equation for the isodoublet fermions in the external
gauge field can be written in the form
\be
i\frac{\partial\Psi}{\partial t}=H_{D} \Psi
\ee
with
\be
H_{D}=-i(\vec{\alpha}\cdot \vec{\nabla}-ig\vec{\alpha}
\cdot \vec{W})\;, \label{hdir}
\ee
and the matrices $\vec{\alpha}$ in a chiral representation are
\be
\vec{\alpha}=\gamma ^{0}\vec{\gamma}=
\left(\begin{array}{cc}
\vec{\sigma} &0\\
0 & -\vec{\sigma} \end{array}\right)\;,
\ee
where $\vec{\sigma}$ are the usual Pauli spin matrices.
We want to show that time-independent Dirac equation
\be
H_D \Psi=E \Psi
\ee
has a normalizable  zero energy solution.

The scalar product is given by
\be
\langle \Psi_1 |\Psi_2 \rangle=\int d^3 x
\Psi_{1}^{\dagger}\Psi_{2}\;.
\ee
One can decompose $\Psi$ into its
left and right chiral components\\
$\Psi=\left(\begin{array}{c} \psi_L\\ \psi_R
\end{array}\right)$. The Dirac equation for the left (right)
component becomes
\be
[\vec{\sigma}\cdot\vec{\nabla}-ig\vec{\sigma}\cdot\vec{W}]
\psi=0\;.
\ee
{}From the ansatz Eq.(\ref{ansatz}) it follows that
\be
\vec{\sigma}\cdot \vec{W}=\frac{W-1}{2gr}
(\vec{n}\cdot\vec{\sigma}\times\vec{\tau})\;.
\ee
Note that the gauge configuration Eq.(\ref{ansatz}) is invariant
under space rotations accompanied by
corresponding gauge transformations.
This implies that the total angular momentum
$\vec{J}=\vec{L}+\vec{S}+\vec{T}$, the sum of orbital angular
momentum $\vec{L}$, spin $\vec{S}=\frac{1}{2}\vec{\sigma}$
and isospin  $\vec{T}=\frac{1}{2}\vec{\tau}$ is conserved.
One can check that $\vec{J}$ commutes with $H_D$.
Therefore one can label the solutions of the Dirac equation
by eigenvalues  $j$ and $m$ of $\vec{J}^2$ and $J_z$.

In the $s-$wave sector ($j=m=0$) the most general form of
a spherically symmetric wave function contains two radial functions
$f(r)$ and $g(r)$ \cite{JR}
\be
\psi(r)=f(r)\chi_1 +g(r)\chi_2
\ee
where $\chi_1$ is the constant spinor satisfying
$\chi_1^\dagger\chi_1=1$ and the "hedgehog" property
\be
(\vec{\sigma}+\vec{\tau})\chi_1=0
\ee
and $\chi_2=\vec{\sigma}\cdot\vec{n}\chi_1$. The explicit form of
$\chi_1$  is $\chi_1,_{\alpha m}=\frac{1}{\sqrt{2}}
\epsilon_{\alpha m}$ $\alpha$ being Lorentz index and
$m$ the isospin index.

Using the "hedgehog" property one can check that
\be
(\vec{n}\cdot\vec{\sigma}\times\vec{\tau})\chi_a=
-2i\epsilon_{ab}\chi_b,\;\;\;{\rm for}\;\;\; a,b=1,2.
\ee
Having this in mind we obtain a system of first order equations
for the radial functions
\bea
\frac{df}{dr}+\frac{1-W}{r}f&=&0 \nn \\
\frac{dg}{dr}+\frac{1+W}{r}g&=&0\;.\label{req}
\eea
These equations can be trivially integrated with the result
\bea
f(r)&=&N \e^{-\int_0^r \frac{1-W}{r}dr}  \nn \\
g(r)&=&0\;,
\eea
where $N$ is a normalization constant.
For $r\to 0$ the function $f_n(r)\to N_n\e^{-b_{n}r^2}$.
For $r\to\infty$

a) for odd $n$ ${f_n(r)}_{r\to\infty}\to\frac{1}{r^2}$
and $f_n$ is normalizable,

b) for even $n$ ${f_n(r)}_{r\to\infty}\to\e^{\frac{c_n}{r}}$,
hence $f_n$ is not normalizable in this case.
\vglue 0.6cm
{\bf\noindent 5. Stability analysis \hfil}
\vglue 0.4cm

In this section we will consider the question of linear
stability of dilatonic spha\-le\-rons.
In order to analyse stability one has to consider the spectrum of the
small (harmonically) time dependent perturbations in the background
solution. Perturbations with negative  energy (complex frequency)
correspond to exponentially growing
modes indicating to the instability of the background solution.

We will consider perturbations of  the type
\bea
\phi_1 \to W(r)+v(r) e^{i\omega t} &\quad&
\phi_2 \to \chi (r)\e^{i\omega t} \cr
\varphi\to\varphi(r)+\psi(r) e^{i\omega t} &\quad&
A_1 \to a(r) e^{i\omega t}
\eea
Calculation of the quadratic action shows that the
$(\chi, a)$ sector decouples from the $(v,\psi)$ sector,
\be
S^{(2)}=S^{(2)}_{(v,\psi)}+S^{(2)}_{(\chi, a)}
\ee
where
\bea
S^{(2)}_{(v,\psi)}&=&\int dr\Bigl(\frac{1}{2} r^2\psi'^2
-\omega^2(\frac{1}{2}r^2\psi^2+e^{2\varphi}v^2) \cr
&+&e^{2\varphi}\Bigl[v'^2-4W'v\psi'+{3W^2-1\over r^2}v^2
+(2W'^2+ {(W^2-1)^2\over r^2})\psi^2 \Bigr] \Bigr)
\eea
is the same as in \cite{LM1} and
\be
S^{(2)}_{(\chi, a)}=\int dr \e^{2\varphi} \Bigl[
\chi'^2+{(W^2-1)\over r^2}\chi^2
+W^{2}a^2+2W'a\chi-2Wa\chi'
-\omega^2(\frac{1}{2}r^2 a^2+\chi^2)
\Bigr]
\label{qact1}
\ee
with $\varphi$ respectively $W$ the background solutions.

The spectrum of the perturbations in the $(v, \psi)$ sector was
already analyzed in \cite{LM1}, \cite{BIZ1}. For those solutions,
which were analysed numerically, $n$ negative modes for the
$n^{\rm th}$ solution were found.

Here we will concentrate on the $(\chi, a)$ sector. The replacement
$\chi\to\e^{-\varphi}\tilde\chi,\; a\to\e^{-\varphi}\tilde{a}$
brings the action Eq.(\ref{qact1}) to the form:
\bea
\tilde S^{(2)}_{(\tilde\chi, \tilde a)}=
\int dr \Bigl[\tilde\chi'^2+
(\varphi''+\varphi'^2+{(W^2-1)\over r^2})\tilde\chi^2
-\omega^2(\frac{1}{2}r^2\tilde{a}^2+\tilde\chi^2)\nn \\
+W^{2}\tilde{a}^2+2W'\tilde{a}\tilde\chi
+2W\varphi'\tilde{a}\tilde\chi-2W\tilde{a}\tilde\chi'
\Bigr]\;. \label{qact2}
\eea
The equation of motion are:
\bea
W(\tilde\chi'-W\tilde{a})-W'\tilde\chi-W\varphi'\tilde\chi&=&
-{\omega^2r^2\over2}\tilde{a}\cr
(\tilde\chi'-W\tilde{a})'-(\varphi''+\varphi'^2+{(W^2-1)\over r^2})
\tilde\chi-W'\tilde{a}+W\varphi'\tilde{a}&=&-{\omega^2}\tilde\chi\;.
\eea
We have performed a numerical analysis of the spectrum for the first
few solutions. We find that $n^{\rm th}$ solution has $n$ negative
modes also in $(\chi, a)$ channel.
In the Table we have collected
our numerical values for the energies $E=\omega^2$
of the negative modes of the first three solutions.
\begin{center}
\begin{tabular}{|l|l|l|}  \hline
$N=1$        &$N=2$        &$N=3$         \\ \hline
$E_1=-0.2098$&$E_1=-0.1477$&$E_1=-0.1238$ \\
             &$E_2=-0.0005$&$E_2=-0.0002$ \\
        &          &$E_3=-6{\cdot}10^{-7}$ \\ \hline
\end{tabular}
\vglue 0.4cm
Tab. Boundstate energies for the $n=1,2,3$ solutions of the
YMD theory,\\
($(\chi, a)$ sector).
\end{center}
Hence altogether we found that the $n^{\rm th}$ solution
of the YMD theory has $2n$ unstable modes with
respect to spherically symmetric perturbations
similarly as for the BMK solutions \cite{LM3}.

\vglue 0.6cm
{\bf\noindent 6. Concluding Remarks \hfil}
\vglue 0.2cm
Static solutions of YMD theory are a kind of sphalerons -
dilatonic sphalerons. In contrast to the electroweak sphaleron
in the YMD system one finds solutions with many zeroes
of the gauge function.

In conformity with the level crossing picture
one finds a normalizable fermion
zero mode in the background of these solutions.

These solutions exist in the Einstein-Yang-Mills-dilaton theory as
well \cite{LM2}. Because of the high mass of the solutions the natural
situation where they could play a role is in the Early Universe.
This question needs further investigation.
%
%
\vglue 0.6cm
{\bf \noindent 7. Acknowledgements \hfil}
\vglue 0.2cm
I am indebted to P. Breitenlohner, P. Forg\'acs and
D. Maison for many stimulating discussions.
I would like to thank N. Straumann for the kind hospitality
at the University of Zurich, where this work was completed.
This work was supported in part by Tomalla Fellowship.
\vglue 0.6cm
{\bf\noindent 8. References \hfil}
\vglue 0.2cm

\end{document}